\documentclass[sigconf]{acmart}
\usepackage{graphicx}
\usepackage{graphics}
\usepackage{amsmath}
\usepackage{algorithm}
\usepackage{algorithmic}
\usepackage{xcolor}
\usepackage{amsmath}
\usepackage{url}
\usepackage{caption,subcaption}

\AtBeginDocument{%
  \providecommand\BibTeX{{%
    \normalfont B\kern-0.5em{\scshape i\kern-0.25em b}\kern-0.8em\TeX}}}

\renewcommand\footnotetextcopyrightpermission[1]{} % removes footnote with conference information in first column
\begin{document}

\title{Snapshot Samplings of the Bitcoin Transaction Network and Analysis of Cryptocurrency Growth}  

\author{Lambert T. Leong}
\affiliation{%
  \institution{Department of Computer Science}
  \institution{University of Hawaii at Manoa}
  \streetaddress{2500 Campus Rd}
  \city{Honolulu}
  \state{Hawaii}
  \postcode{96822}
  \country{USA}}
\email{lambert3@hawaii.edu}

\begin{abstract}
The purpose of this work was to perform a network analysis on the rapidly
growing bitcoin transaction network.  Using a web-socket API, we collected data
on all transactions occurring during a six hour window.  Sender and receiver
addresses as well as the amount of bitcoin exchanged were record.  Graphs were
generated, using R and Gephi, in which nodes represent addresses and edges
represent the exchange of bitcoin.  The six hour data set was subsetted into a
one and two hour sampling snapshot of the network.  We performed comparisons
and analysis on all subsets of the data in an effort to determine the minimum
sampling length that represented the network as a whole.  Our results suggest
that the six hour sampling was the minimum limit with respect to sampling time
needed to accurately characterize the bitcoin transaction network.Anonymity is a
desired feature of the blockchain and bitcoin network however, it limited us in
our analysis and conclusions we drew from our results were mostly inferred.
Future work is needed and being done to gather more comprehensive data so that
the bitcoin transaction network can be better analyzed. 
\end{abstract}

\maketitle

\section{Introduction}
\label{sec:intro}

The invention of the internet has helped to provide a new platform for the
exchange of money by providing the framework to connect individuals from all
around the world.  Most monetary exchanges over the internet occur with the
help of third party entities such as banks or credit card companies.  These third
party entities often have oversight over all transactions and are responsible
for verifying the integrity of transactions~\cite{money}. Exchange protocols, which
require a verifier, have become the conventional method of exchanging money
over the internet.   While these conventional exchange methods account for a good
amount of the daily transactions around the world, there are a few draw backs
that present problems to many users.  Third party verifiers constantly receive
a high volume of transactions which need to be verified before money is
permanently moved from one individual to another.  Different companies and
banks have different methods of prioritizing the order in which transactions
are verified.  A high number of transactions verifications by a small number of
third parity entities leads to a bottle-necking effect which can leave
recipients waiting for payments to clear and be received.  In an effort
to reduce the volume of transactions many third party verifiers require minimum
transactions amounts and, in some cases, a small transaction fee.  While this
may help to mitigate the volume of transactions that need to be verified, it
limits the financial freedom of those who participate in conventional exchange
methods.  

Bitcoin (BTC) is a cryptocurrency that offered a solution to the shortcomings present
in conventional exchange methods.  Invented by Satoshi Nakamoto, bitcoin is a
peer-to-peer payment system in which payments are validated by math rather than
trust~\cite{bitcoin}.  In other words, instead of trusting a potentially
nefarious third party entity, transactions are verified by a proof of work
algorithm which involves block
hashing~\cite{bitcoin,anonymityanalysis,qa_btc_graph}.  A public ledger, known
as the blockchain, is created as a result of the hashing algorithm and it
contains every transaction in the history of the network.  New blocks are
hashed about every 10 minutes and transactions are verified and added to the
ledger indicating a successful transfer of bitcoin from one individual or
entity to another~\cite{bitcoin}. Anonymity is preserved by keeping user
addresses separate and abstracted away from personal information.  In addition,
there are some instances where a particular user's wallet can generate new
sending and receiving addresses to further maintain a level of
anonymity~\cite{bitcoin, andreas}.  

Since its genesis, bitcoin has gained popularity and traction.  As a result,
the number of bitcoin addresses and the number of transactions have continued
to increase at an exponential rate as seen in Figures~\ref{fig:btc_info}.  In
this paper, we analyze different snapshots of the growing bitcoin network.
Since the bitcoin network is currently large and continuously growing, we are
interested in investigating the appropriate network sampling time frame.  Network
and graph analysis have been performed on the bitcoin network however, its
rapid growth may indicate a need for a better way of characterizing the
network at a specific point in time~\cite{graph_ana_btc}. Determining a
sampling criterion would allow analysis to be done on a smaller sub-graph from
which, conclusions can be drawn and extrapolated to the network as a whole.
Running a network analysis on a representative sub-graph would be more
efficient and computationally favorable than the whole, growing, bitcoin
network.

\iftrue
\begin{figure}[h!]
    \centering

\begin{subfigure}{0.45\textwidth}
    \centering %
    \includegraphics[width=.9\textwidth]{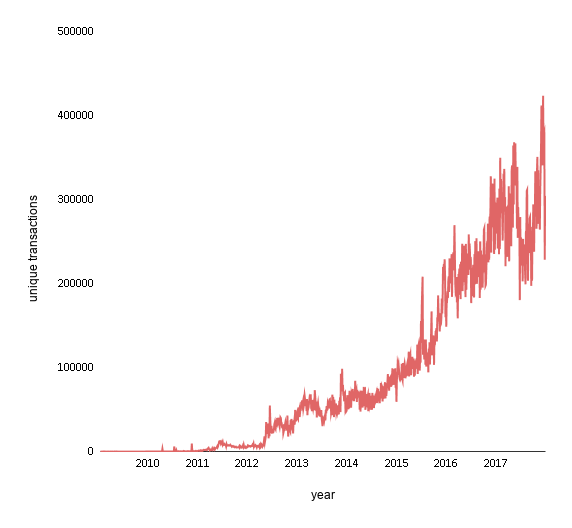}
    \caption{Number of transactions over time grows as an exponential function}
    \label{fig:trans}
\end{subfigure}
%\qquad
\begin{subfigure}{0.45\textwidth}
    \centering %
    \includegraphics[width=.9\textwidth]{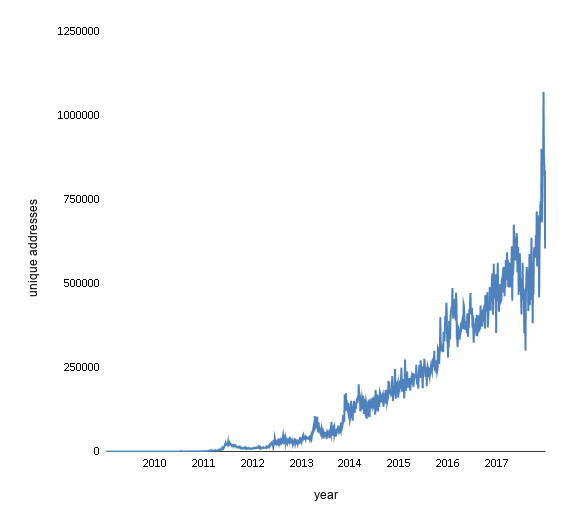}
    \caption{Number of unique Bitcoin addresses growth over time}
    \label{fig:address}
\end{subfigure}

    \caption{Growth in the number of transactions and unique addresses imply the growth in bitcoin users and popularity}
    \label{fig:btc_info}
\end{figure}
\fi

The remainder of the papers is as follows: in Section~\ref{sec:methods} we
describe how data is collected, how we generate graphs from that data, and how
we use specific tools to analyses the network. In Section~\ref{sec:results} we
report our finding at different sampling intervals.  Lastly, we discuss our
findings and conclusion in Section~\ref{sec:conclusion}.

\section{ Methods}
\label{sec:methods}

BTCs are constantly being exchanged and moving throughout the network and
as a result the numbers of transactions are ever growing.  The number of users
and addresses are also constantly increasing.   Every transaction involving
BTC is public and every transaction since BTCs inception in 2009 is
recorded on a public ledger.  This is a unique feature of BTC and every
node participating in the network has a copy of this public ledger.  This
ledger is constantly being updated as new transactions are confirmed.  Updates
occur on every ledger on every node participating in the network and thus every
node is in consensus with each other on the current status of the ledger.  

\subsection{Data Collection}

While we could have joined the blockchain network and downloaded a copy of the
public ledger to build our network graph, we felt that it would not be an
efficient method of evaluating the current status of the BTC network.  As a
result, we decided to utilize a web socket API from BlockChain.Info.  This API
allowed us to stream live transactions as they were on their way to be
confirmed.  Each transaction came in as a JavaScript Object Notation (JSON)
file and a script was written to extract and parse out the sender address,
receiver address, and the BTC amount variables.  These variables were stored as
a comma separated file (.csv).  The web socket ran for a total of six hours and
separate csv files were generated at the one hour, two hour, and six hour time
points; hence the snapshots.  One of our objectives was to investigate possible
changes in network characteristics with different sample sizes and durations.
In other words, we are interested to see if any network metrics change as the
network grows within our sampling snapshots.  It is important to note that the
web socket performed a continuous data collection and was not interrupted when
the one hour and two hour samples were taken.  The resulting one hour and two
hour samples are sub sets of the six hour sample with the one hour sample also
being a subset of the two hour sample.

\subsection{Graph Generation}

The csv files for the one, two, and six hour samplings were imported into
Gephi~\cite{gephi} to generate a graph of the sampled BTC network.  The csv
files were imported as edge list where the sending address corresponded to the
``source'' and the receiving address corresponded to the ``target''.  The amount
was converted from Satoshis to BTCs, 1 BTC = $1$x$10^{-8}$ Satoshi, and stored as a
float, an edge attribute.  Graph files were exported as .graphml for further
data processing.

\subsection{Network Analysis}

Graph metrics were analyzed with R utilizing the igraph, poweRlaw, and linkcomm
packages~\cite{igraph}.  Transitivity, average degree, average distance,
reciprocity, degree distribution, and maximal cliques were measured for each
snapshot graph and compared to each other.  The largest connected components
were also extracted from each of the three sample graphs and metrics were
computed with igraph.  The six hour connected graph was further evaluated for
the presence of communities.  Using the linkcomm package and the
getlinkcommunities, by Alex T.  Kalinka, we were able to generate community's
nodes that linked to every node of a particular community~\cite{igraph}.
Investigating community structures in the giant connected component may reveal
BTC user financial behavior; in particular who exchanges money with whom.

\section{Results}
\label{sec:results}

There was a positive correlation between the sampling time and the amount of
observed transactions.  As mentioned previously in Section~\ref{sec:intro}, the
numbers of transactions are increasing at an exponential rate.  We plotted the
number of edges, obtained in our sampling, over the duration of six hours.
Transactions are indicated by an edge in the resulting graph of the networks
sampled for one, two, and six hours.  The increase or growth of the network
resulted in an increase in both the number of transactions, indicated by the number
of edges, and the number of active addresses, indicated by the
number of nodes.  These results are shown in Figure~\ref{fig:sampling}.  

\iftrue
\begin{figure}[h!]
    \centering
    \includegraphics[width=.4\textwidth]{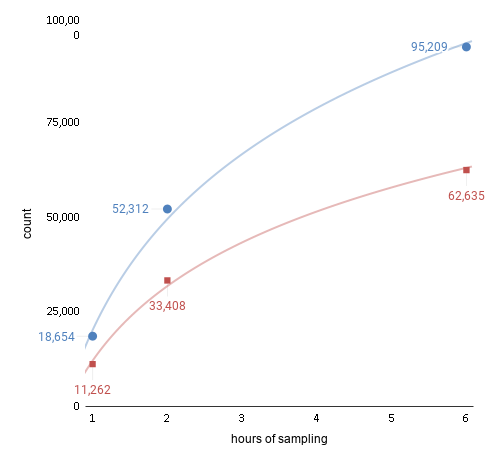}
	\caption{As we take longer samples, the network grows.  Nodes \textcolor{blue}{(BLUE)} indicate new
	addresses, and edges \textcolor{red}{(RED)} indicates new transactions.  R$^{2}$ values for best
	fit lines for nodes and edges are 0.997 and 0.996 respectively}
	\label{fig:sampling}
\end{figure}
\fi

The trend line would indicate that the number of transactions are growing at a
logarithmic pace rather than an exponential one.  R$^{2}$ values for
logarithmic best fit lines for nodes and edges are 0.997 and 0.996 respectively
This is contradictory to what was expected however, this discrepancy
may be due to the  amount of data and sample size.  Blockchain.info
reports an exponential growth in the number of transactions over the whole life
span of BTC~\cite{blockchain.com}.  Our sampling window could be the limiting
factor and perhaps, we would start to see an exponential growth trend for a
longer sampling duration.  

The number of transactions since the inception of BTC has an overall general
exponential growth trend but the transaction rate seems to fluctuate if
observed form the perspective of weeks and months.  There is a level of
volatility when it comes to the prices or exchange rate of BTC~\cite{econ}.
Price volatility could correlate to these fluctuations and periods of lower
transaction volumes.  A six hour sampling window does not appear to be a long
enough time frame to model the overall network growth.  Since transactions
rates can fluctuate for periods of time ranging from days to months, it is best
to look at the entire transaction history in order to draw conclusions about
the overall growth rate of transactions.  A smaller sampling window, such as the 6
hours for which we took our samples, may be useful when analyzing how the
network is growing in the near or short term.  

\subsection{Snapshot Network Analysis Results and Comparisons}

Network analysis was performed on all three graphs with the purpose of
investigating changes as the time and the size of the sample network increased
as well as for comparison to previously reported findings.  Analyzing the
difference in resulting graph metrics would help in determining an appropriate
sampling window that would yield a sample graph that is representative of the
network as a whole. Table~\ref{tab:data_count} contains the metrics calculated for all three
network graphs.

\iftrue
\begin{table}[h]
\centering
\caption{Graph metrics for one, two, and six our sampling snapshots calculated using igraph and R}
\resizebox{.5\textwidth}{!}{
%\begin{tabular}{|l|c|c|c|}
\begin{tabular}{c|lcr}
\toprule
Sampling Duration & 1 Hour & 2 Hour & 6 Hour \\ 
\hline
\hline
Nodes & 18654 & 52312 & 95209 \\ 
Edges & 11262 & 33408 & 62635 \\ 
Diameter & 12 & 19 & 39 \\ 
Max Cliques & 3 & 3 & 3 \\ 
Dyads & 10943 & 32512 & 61122 \\ 
Tryads & 2 & 8 & 34 \\ 
\hline
Reciprocity & 0.00 & 0.00 & 0.00 \\ 
Transitivity (global) & 0.00 & 0.00 & 0.00 \\ 
Transitivity (average) & 0.00 & 0.00 & 0.00 \\
Mean Degree & 1.21 & 1.28 & 1.32 \\ 
Mean Distance & 1.31 & 1.90 & 5.37 \\ 
\bottomrule
\end{tabular}
}
\label{tab:data_count}
\end{table}
\fi

\iftrue
\begin{figure*}[h!]
    \centering

\begin{subfigure}{0.3\textwidth}
    \centering %
    \includegraphics[width=.99\textwidth]{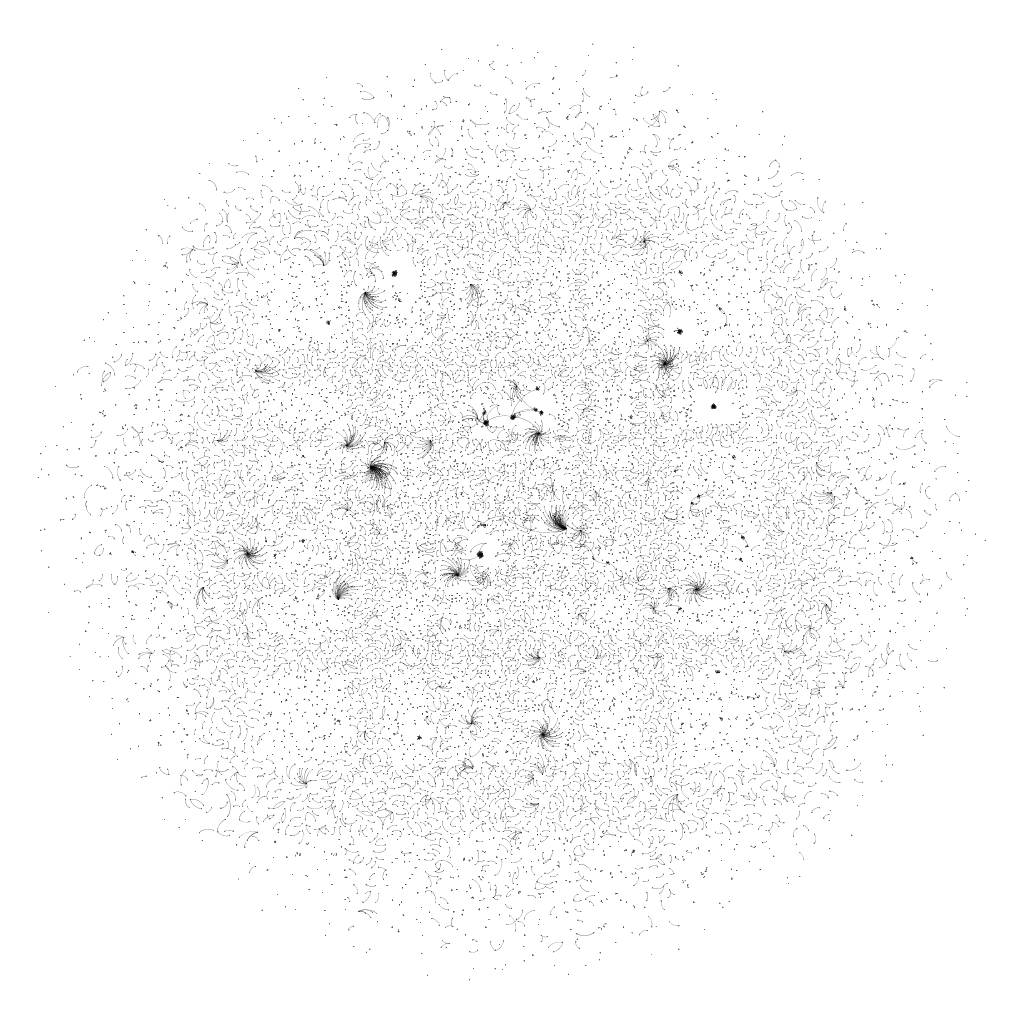}
    \caption{Graph sampling after one hour}
    \label{fig:1hour0}
\end{subfigure}
\hfill
\begin{subfigure}{0.3\textwidth}
    \centering %
    \includegraphics[width=.99\textwidth]{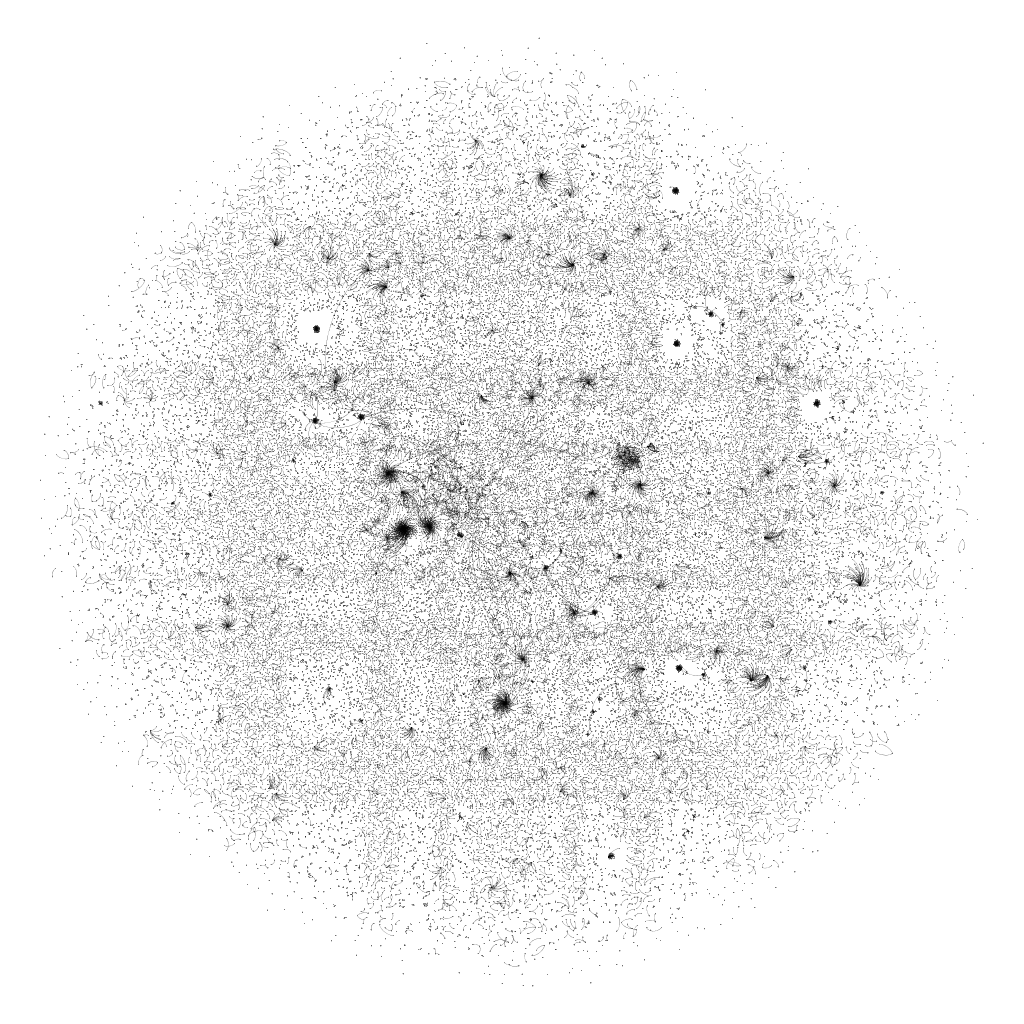}
    \caption{Graph sampling after two hours}
    \label{fig:2hour0}
\end{subfigure}
\hfill
\begin{subfigure}{0.3\textwidth}
    \centering %
    \includegraphics[width=.99\textwidth]{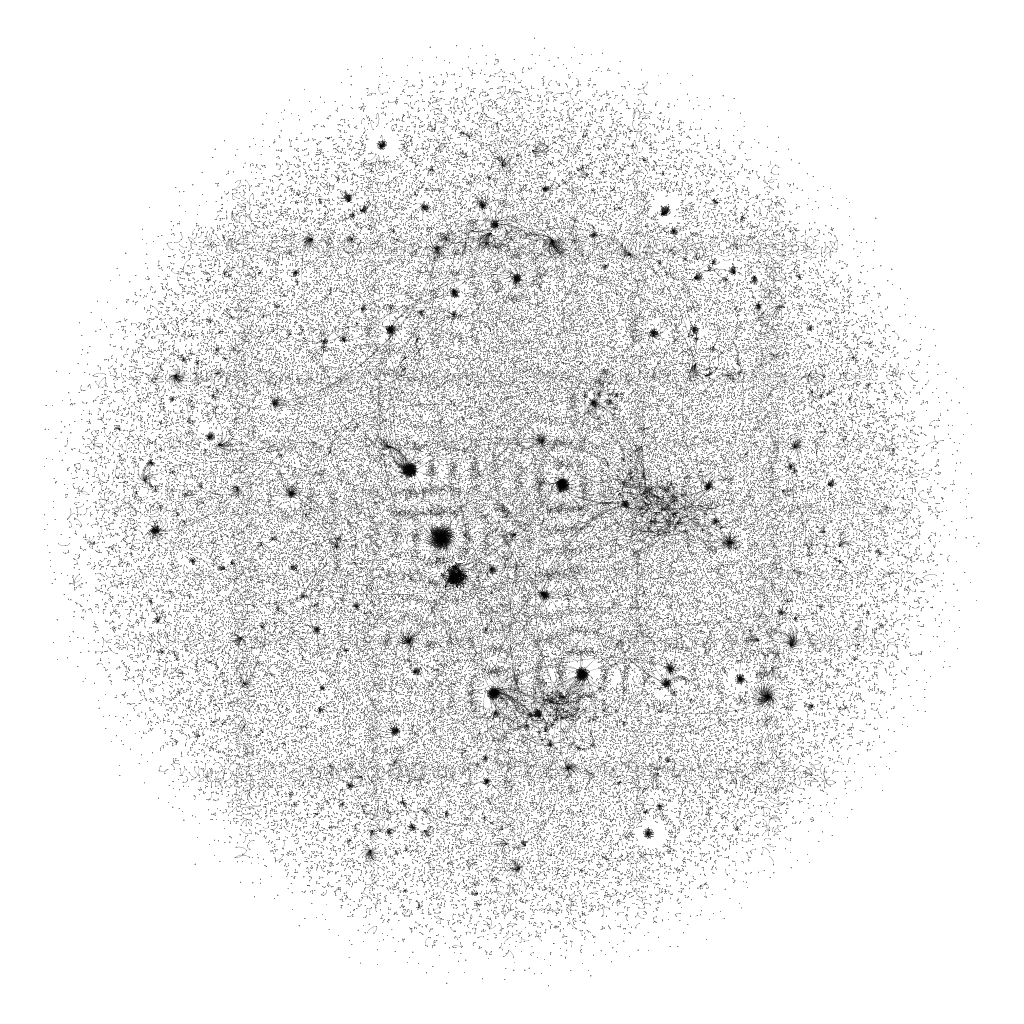}
    \caption{Graph sampling after six hours}
    \label{fig:6hour0}
\end{subfigure}

    \caption{Visualizations of one, two, and six hour sample graphs, from left to right.  Graph density increases as collection time increases}
    \label{fig:graphs}
\end{figure*}
\fi

\begin{figure*}[h!]
  \centering \includegraphics[width=\textwidth, height=0.25\textheight]{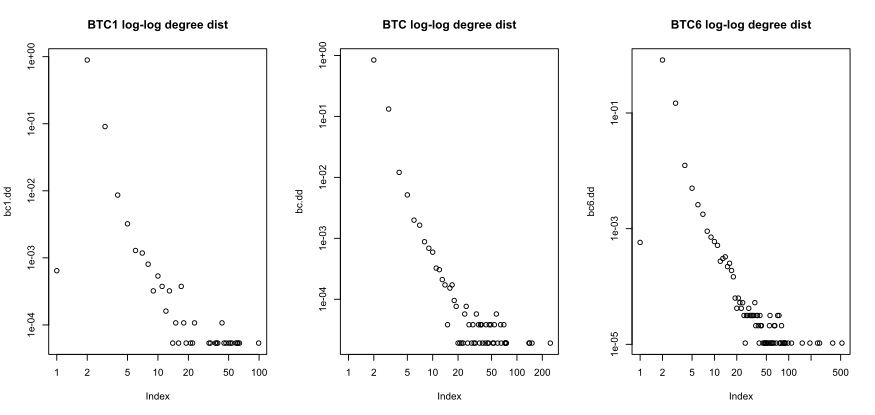}
  \caption{Log-Log degree distribution of one, two, and six hour networks, from left to right. }
  \label{fig:powerlaw1}
\end{figure*}

Across all three sample graphs, reciprocity, global transitivity, and mean
degree were consistent.  These metrics changed only slightly as the
graph grew over the 6 hours.  Transitivity was low for all three graphs, which
was to be expected.  Transitivity would increase when, for example, someone
sent BTC to two separate people and those two people exchanged BTC with
each other.  This is not a likely event and this is why we expected it to be
low.  All three graphs had a low mean degree with the largest mean degree value
being 1.32 for the six hour sample graph.  A low mean degree would help to
explain a low transitivity since a low mean degree indicates that the majority of
the nodes in a graph have only one edge.  Nodes need a minimum of two edges for
transitivity.  The mean degree values for each graph indicate that most
addresses execute only one transaction.  The number of dyads implies that the
graph is primarily composed of pairs of nodes that share only one edge.  For
each sample graph, dyads account for the majority of nodes in the network.
Triads make up a significantly lesser portion of the graph.  We did not expect
to see many triads or cliques during our sampling.  In the
short term, it is unlikely that triads form because it would mean that some
money comes back to the original sender implying that some portion of that
transaction should not have been sent out in the first place.  However, goods
or services could have been exchanged among the triad which would explain its
formation.  The network is built only on BTC transactions and thus we do not
know the exact motive behind each transaction.  A longer sampling would
probably have led to more triads and possibly greater maximal cliques.
However, it is just as likely that the number of pairs of nodes would have
increased at a similar rate.    

Reciprocity was also low.  This was expected for this type of network.  It is
unlikely that a node would reciprocate an edge because that would be like
person A sending 1 BTC to person B in the first transaction then person B
sending back, to person A, a given amount of BTC in a second transaction.
The first and second transaction can be combined into one and the net amount,
the difference in the amount between the first and second transaction, should
be sent to the appropriate person which would eliminate an edge in one
direction.  The reciprocity value is small for all graphs but it is not zero
which means that there are instance of reciprocating edges.  The instances
where a user sends the wrong amount to another user who, is kind enough to
correct the transaction and send it back to the person it originated from may
be an instance for reciprocation.  Returning goods and getting a refund on
something bought with BTC is another instance that may warrant
reciprocation.  Reciprocation can also be the result of gambling sites where,
gamblers gamble in BTC and earn their winnings in BTC too.  These
instances are rare and small in comparison to the vast number of transactions
constantly being received.  The rarity of reciprocity is reflected in the zero value
obtained for all sample graphs.   

Visual representations of each samples size graph, density, and connections are
show in Figure~\ref{fig:graphs}.  The density of the graph increase as sampling
time increases.  Dark black spots are star like structures consisting of
clusters of nods around a single node with a high degree.   As time goes on,
more nodes form edges to these central nodes which increase the size of the
black spots.  However, it does not appear that links are formed from one high
degree node to another.  Nodes with high degree seem to increase their degree
as more nodes join the network which exhibits characteristics of preferential
attachment and implies a possible scale free network~\cite{barabasi, rich}.  

\begin{figure*}[h!]
  \centering \includegraphics[width=\textwidth, height=0.25\textheight]{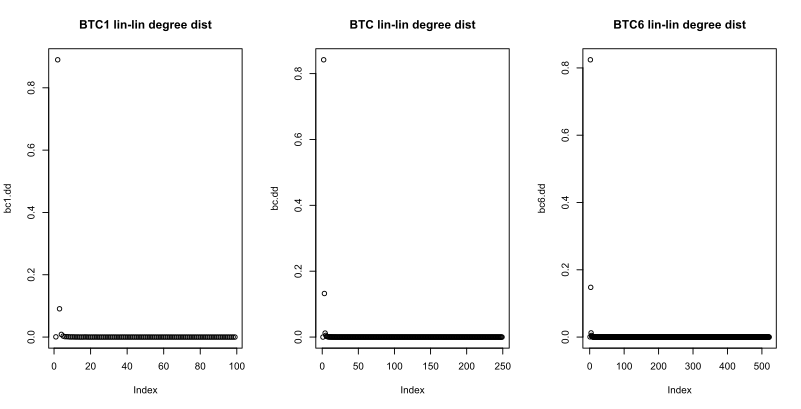}
  \caption{Lin-Lin degree distribution of one, two, and six hour networks, from left to right.}
  \label{fig:powerlaw2}
\end{figure*}

\begin{figure*}[h!]
  \centering \includegraphics[width=\textwidth, height=0.25\textheight]{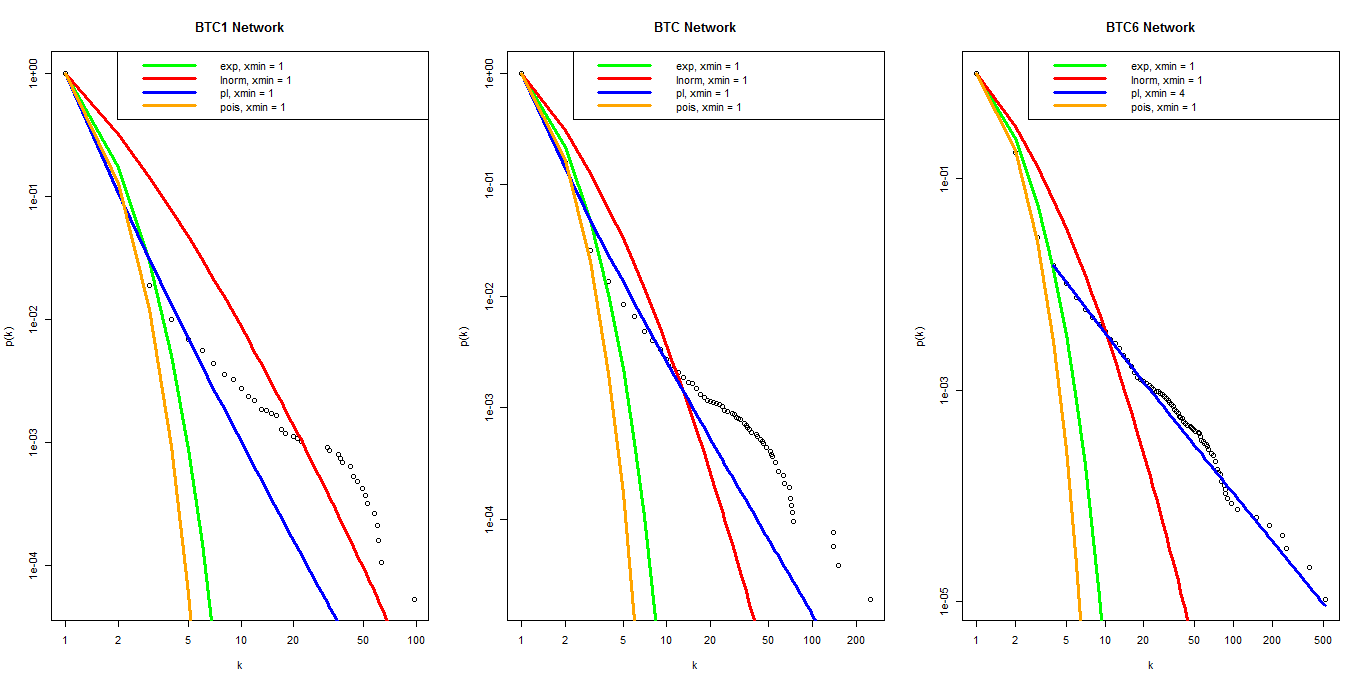}
  \caption{Plot of one, two, and six hour network degree distributions with exponential(Red), log normal(Green), power-law(Blue), and Poisson(Orange) best fit lines, from left to right.  }
  \label{fig:degree_dist}
\end{figure*}

Scale free networks are characterized by having a power-law degree
distribution.  We investigated the degree distribution in
Figure~\ref{fig:powerlaw2}. Here we see that the degree distribution for all
sample networks is heavily tailed to the right which is characteristic of
power-law distributions.    

The degree distributions for each graph was plotted in conjunction with
exponential, log normal, power law, and Poisson best fit lines.  Results in
Figure~\ref{fig:degree_dist} would indicate that the one hour and two hour
graphs do not fit any particular trend.  However, when $X$min was set at 4, the
six hour distribution seemed to fit the power law distribution.
Kolmogorov-Smirnov test results in a p value of 0.46 which indicates that we
cannot reject the hypothesis that the six hour distribution fits a power
law~\cite{igraph}.  

Evidence of a power law distribution were seen in our largest graph which was
sampled for the longest period of time.  It would appear that smaller samplings
of the network do not resemble a scale-free network however, the largest
sampling started to show evidence of a power law distribution which is
indicative of a scale-free network.  It is likely that a longer sampling window
would yield a larger graph in which a power law is more apparent.  

Although the graphs are mainly composed of pairs of nodes, we were interested
in the presence of communities.  We extracted the largest connected component
from the sample graphs.  This left us with graphs that excluded the many
isolated pairs of nodes.  Metrics were calculated and are shown in Table~\ref{tab:data_count2}.

\begin{table}[h]
\centering
\caption{Graph metrics for giant component subgraph calculated using igrap}
\resizebox{.5\textwidth}{!}{
%\begin{tabular}{|l|c|c|c|}
\begin{tabular}{c|lcr}
\toprule
\begin{tabular}[c]{@{}c@{}}Connect Component\\ Sampling Duration\end{tabular} & 1 Hour & 2 Hour & 6 Hour \\ \hline
\hline
Nodes & 238 & 3028 & 9052 \\ 
Edges & 243 & 3193 & 9698 \\ 
Diameter & 4 & 19 & 39 \\ 
Max Cliques & 2 & 3 & 3 \\ 
Dyads & 2 & 3172 & 9582 \\ 
Tryads & 0 & 3 & 20 \\ 
\hline
Reciprocity & 0.00 & 0.00 & 0.00 \\ 
Transitivity (global) & 0.00 & 0.00 & 0.00 \\ 
Transitivity (average) & 0.00 & 0.00 & 0.01 \\
Mean Degree & 2.04 & 2.11 & 2.14 \\ 
Mean Distance & 1.29 & 3.04 & 6.67 \\
\bottomrule
\end{tabular}
}
\label{tab:data_count2}
\end{table}

\iftrue
\begin{figure}[h!]
    \centering

\begin{subfigure}{0.45\textwidth}
    \centering %
    \includegraphics[width=1.\textwidth]{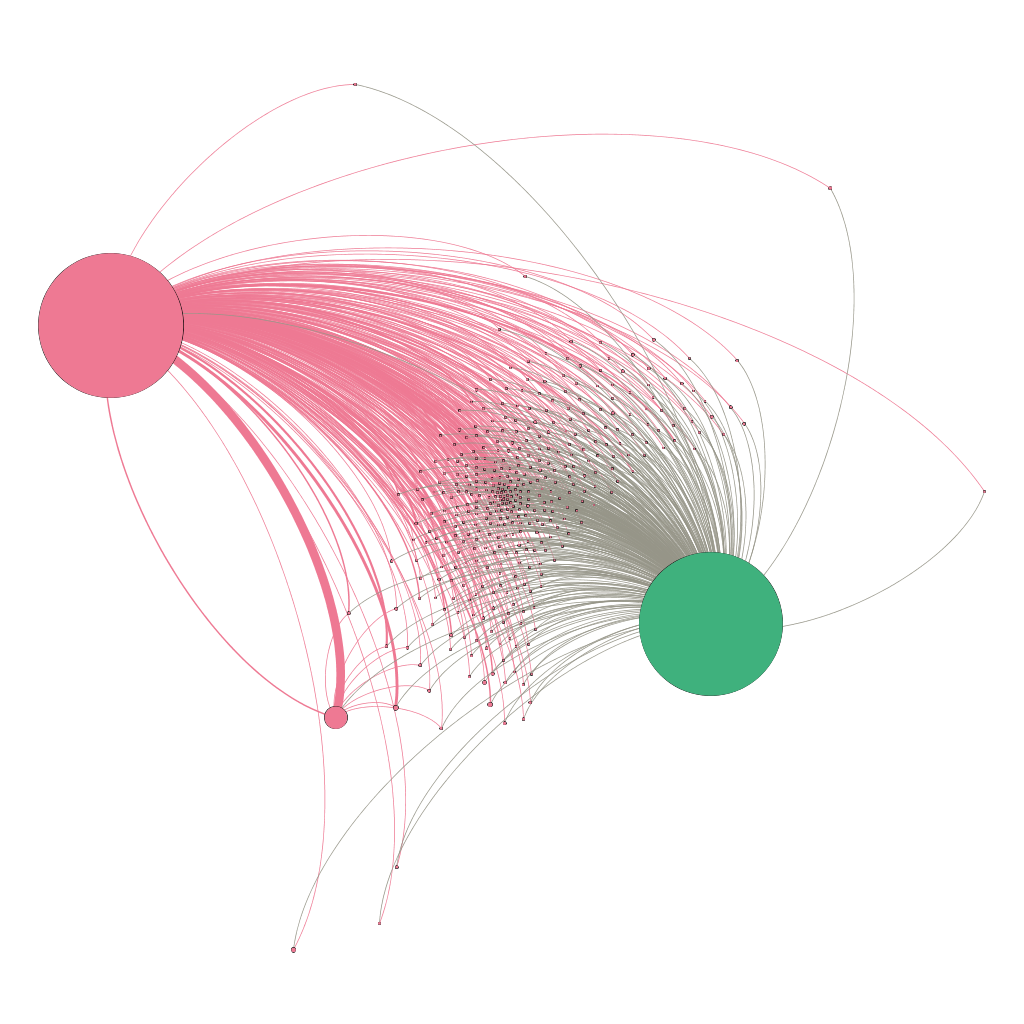}
    \caption{Large community in which many nodes share only one edge with one of the two main nodes}
    \label{fig:1hour}
\end{subfigure}
%\qquad
\begin{subfigure}{0.45\textwidth}
    \centering %
    \includegraphics[width=1.\textwidth]{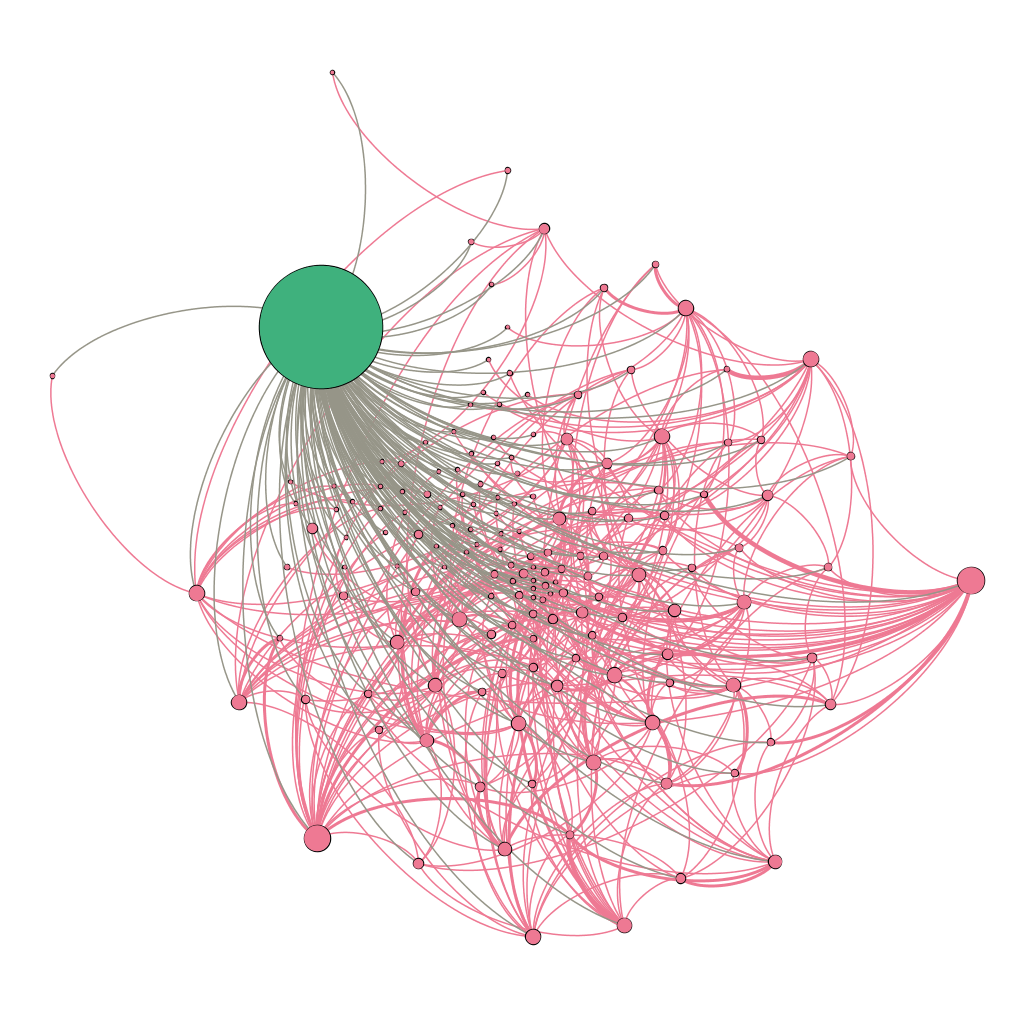}
    \caption{Large community which has a more decentralized shape}
    \label{fig:2hour}
\end{subfigure}

    \caption{Two of the largest communities from the six hour sample.  Nodes are sized by the number of edges.  Bigger nodes have more edges.  Green colors indicate that a higher in degree (receiving BTC) while pink represents a higher out degree (sending BTC)}
	\label{fig:linkgraphs}
\end{figure}
\fi

\subsection{Analysis of Six Hour Snapshot Giant Components}

Directedness was not taken into account when extracting the giant components
from each graph.  Each sub graph is significantly smaller than its original
graph implying again that the graphs are primarily made up of isolated pairs of
nodes.  This means that, during the sampling window, addresses only sent one
transaction.  For the giant component subgraphs, the metrics were not
consistent across all giant component subgraphs, as seen in the original
graphs.  

Our analysis showed that the six hour graph showed to be the best
representative sample network.  Therefore, we chose to run community detection
on the six hour giant component subgraph.  The resulting graphs can be viewed
in the appendix.  Figures~\ref{fig:linkgraphs} are sub graphs of the graph, Figure~\ref{fig:bigcomm}, in the appendix.   

Both graphs represent the two biggest communities contained in the six hour
giant component graphs and they have slightly different structure.  Figure~\ref{fig:1hour}
illustrates a community in which many nodes share edges with one central node.
This creates a star like structure seen in most of the other communities.  The
second community in Figure~\ref{fig:2hour} does not contain a single node with a high degree.
Most of the nodes have the same degree and yet they form a community.  This
would imply that the nodes are more interconnected to each other, as there is
no distinct hub or authority.  Figure~\ref{fig:2hour} represents more of a decentralized
community.  

Community structures provide clues as to what types of individual or entity may
be contained within a network.  For communities that resemble that of Figure~\ref{fig:1hour},
it is possible that the high degree node may be a type of vendor or a spender
depending whether or not there is a high in-degree or out-degree. A high
in-degree would mean that one address is receiving a lot of transactions while
a high out-degree would mean that an address is paying a lot of other
individuals.  Figure~\ref{fig:2hour} resembles a different type of community.  It is
difficult to draw conclusions about the nature of the nodes in this community.
Gambling is one possible instance in which many transactions occur among a
community of individuals.  However, it is also likely this community is a bunch
of nodes doing business together.  Given that it was a six hour window, it is
unlikely that business would execute that many transactions with each other.
Further analysis is needed to determine the nature of the individuals but
community detection is a good starting point.       

\section{Conclusion}
\label{sec:conclusion}

The six hour sample graph seemed to provide a sample graph that was most
representative of the reported nature of the bitcoin transaction network~\cite{koshy2014}.
Modeling the network growth with the one, two and six hour subgraphs did not
seem to be a reliable means of evaluating long term growth of the network.
Growth fluctuations are experienced in the short term and thus any short term
network growth sampling may not be indicative of the network's overall growth. 

Sample graphs yielded metrics that we expected to see.  Our six hour sample
graph showed evidence of a possible power law degree distribution which suggest
that we cannot dismiss the possibility that our network is scale-free.  The
emergence of such correlation was only seen in the six hour graph and not in
the two hour or one hour graphs.  Further investigations should consider a
longer sampling window in order to get a better snap shot of the network.  A
longer sample window may be beneficial and provide a more definitive
distribution.  

\subsection{Possible Future Work}
A deeper investigation into the owners of each address would aid in a more
comprehensive analysis of the bitcoin transaction network.  Certain security
feature most likely lead to some level of skewness in the data set.  The
feature which allows users to change the sending and receiving address is a
powerful security feature but it could lead to a misrepresentation of the
amount of actual bitcoin users participating in the network.  Since all
transactions are kept on a public ledger, addresses associated with a
transaction involving a lot of bitcoin may become targets of malicious attacks.
Some bitcoin wallets allow users to generate new public addresses in order to
maintain a level of anonymity on the network.  This feature may explain the
large number of pairs of nodes in our network samples.  It could be that
individuals change their address for every transaction and a new address is
used for subsequent transactions.  Situations are present where an address
cannot be changed or it would be inconvenient and this is why we see nodes with
high degrees despite the fact that an address can be changed.  For example, if
a vendor accepts bitcoins as payment it would be inconvenient for the vendor
and patrons if the vendor constantly changed addresses because patrons,
especially ``regulars'', would have to constantly check that the address is
current and that they are sending money to the right address.  

Individuals are allowed to have multiple bitcoin wallets just as someone may
have multiple bank accounts or credit cards.  Multiple wallets per individual
would lead to a similar phenomenon to generating new addresses however; one
more thing needs to be considered.  Money can be sent from one wallet to
another wallet but one individual may have ownership to both wallets; it is
like transferring funds between two bank accounts you own.  To acknowledge this
transfer the, the network needs to verify it and thus it would show up as a
transaction in the network. 

Community detection may help to investigate ownership of a particular
address because multiple addresses in a particular community may belong to one
individual.  Combining these addresses would lead to a better resolution of the
network.  The way it stands now, our network primarily resembles the movement
of bitcoin.  Building a transaction network in which the nodes represent a
single individual or entity rather than a single address would result in a
network that would resemble how people spend bitcoin.  Separating the identity
from the bitcoin address was in the original design of the network to protect
individual 

\section{NOTES}
This paper was written in early 2017 but not put on arxiv until 2020. Please keep the written date in mind when considering analysis, content, state of cryptocurrencies, and etc.

%\noindent \textbf{Reproducibility:} Source code and instructions for deployment are available at  - {\color{blue}\url{https://lambertleong.com/projects}}.

%\section{Appendix}
\iftrue
\begin{figure*}[h!]
    \centering
    \includegraphics[width=1.1\textwidth]{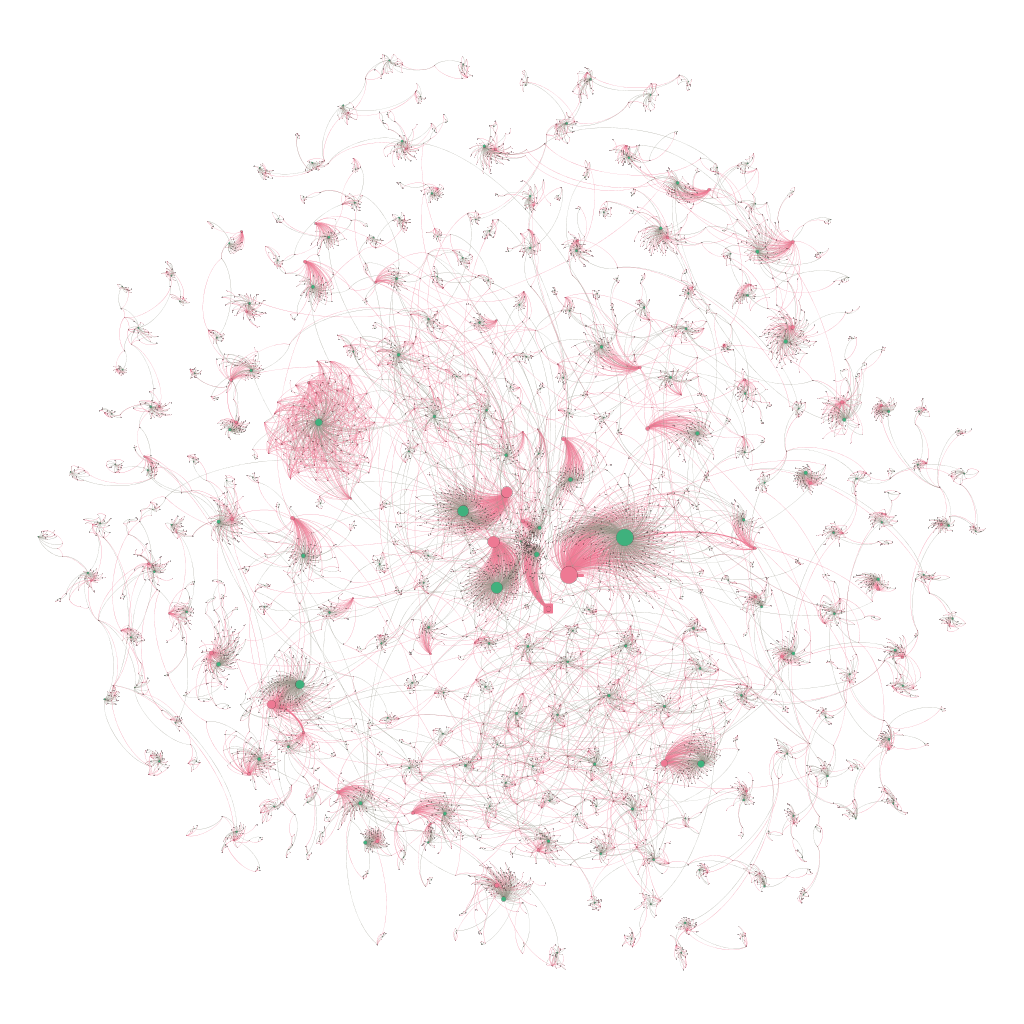}
\caption{Communities from the six hour sample.  Nodes are sized by the number
of edges.  Bigger nodes have more edges.  Green colors indicate that a higher
in degree (recieving BTC) while pink represents a higher out degree (sending
BTC)}
        \label{fig:bigcomm}
\end{figure*}
\fi

\bibliographystyle{acmart}
\bibliography{acmart}
\end{document}